\documentclass[prl,twocolumn,superscriptaddress,showpacs]{revtex4}
\usepackage{hyperref}
\usepackage{amssymb, amsmath}
\usepackage[dvips]{graphicx}
\usepackage{latexsym}
\usepackage{verbatim}

\begin{document}
\title{Superconducting atomic contacts under microwave irradiation}
\date{\today}
\author{M. Chauvin}
\author{P. vom Stein}
\author{H. Pothier}
\author{P. Joyez}
\affiliation{Quantronics Group, Service de Physique de l'\'Etat Condens\'{e}
(CNRS URA 2464), DSM/DRECAM, CEA-Saclay, 91191 Gif-sur-Yvette Cedex, France}
\author{M. E. Huber}
\affiliation{Department of Physics, University of Colorado at Denver, Denver,
Colorado 80204, USA }
\author{D. Esteve}
\author{C. Urbina}
\affiliation{Quantronics Group, Service de Physique de l'\'Etat Condens\'{e}
(CNRS URA 2464), DSM/DRECAM, CEA-Saclay, 91191 Gif-sur-Yvette Cedex, France}

\begin{abstract}
We have measured the effect of microwave irradiation on the dc current-voltage
characteristics of superconducting atomic contacts. The interaction of the
external field with the ac supercurrents leads to replicas of the supercurrent
peak, the well known Shapiro resonances. The observation of supplementary
fractional resonances for contacts containing highly transmitting conduction
channels reveals their non-sinusoidal current-phase relation. The resonances
sit on a background current which is itself deeply modified, as a result of
photon assisted multiple Andreev reflections. The results provide firm support
for the full quantum theory of transport between two superconductors based on
the concept of Andreev bound states.
\end{abstract}  \maketitle

A thorough and unifying view of superconducting electrical transport
emerged in the last fifteen years in the framework of mesoscopic
superconductivity. It is based on the concept of Andreev reflection,
the microscopic process which couples the dynamics of electrons and
holes. In particular, the Josephson currents flowing between two
weakly coupled superconductors are described as arising from Andreev
bound states forming in each conduction channel of the coupling
structure \cite{Furusaki-Tsukada}. The theory predicts the
time-dependent current through a voltage biased short single
conduction channel of arbitrary transmission probability \cite{MAR},
and in particular the interplay between these ac Josephson currents
and a microwave\ external signal \cite{Cuevas2002}. Although ac
supercurrents have been known and detected since the early days of
Josephson circuits \cite{OlddaysACcurrents}, this modern view has
the advantage of being completely general as it applies to all
possible coupling structures, which can always be  decomposed, at
least in principle, into a set of independent channels. In this
Letter we present a test of these predictions carried out on single
atom contacts between two superconducting electrodes
\cite{reviewALR2003}. These contacts are model systems which allow
for a direct comparison of theory and experiment, as one can vary
and measure \cite{scheer} their ``mesoscopic PIN'', \textit{i.e.}
the set of transmission coefficients $\left\{ \tau _{i}\right\} $
characterizing their conduction channels.

 In a single short channel of transmission $\tau $
between two reservoirs with superconducting phase difference $\delta$, two
Andreev states contribute to the Josephson coupling. They have energies $E_{\pm
}\left( \delta ,\tau \right) =\pm \Delta \sqrt{1-\tau \sin ^{2}(\delta /2)}$
lying inside the superconducting gap extending from -$\Delta $ to $\Delta $.
At a given $\delta $ the two states carry opposite currents $I_{\pm }\left(
\delta ,\tau \right) =\left( 1 /\varphi _{0}\right)
\partial E_{\pm }/\partial \delta $, where
$\varphi _{0}=\hbar/2e$ is the reduced flux quantum, and the net current
through the channel results from an imbalance of their occupation numbers. For
a perfect voltage bias $V$ the phase evolves in time according to $\delta
\left( t\right) =\omega _{J}t$ where $\omega _{J}= V/\varphi _{0}$ is the
Josephson frequency. Because the current-phase relation of each state is
periodic, there are ac supercurrents at the Josephson frequency and all its
harmonics, and the current can be written as a Fourier series $I(V,\tau,
t)=\sum\limits_{m} I_{m}(V,\tau )\mathrm{e}^{im\omega _{J}t}$ \cite{Fourier}.
 The sine components arise from the adiabatic evolution of the system on the
ground Andreev state, whereas the cosine components originate in non-adiabatic
(Landau-Zener) transitions between the levels induced by the dynamics of the
phase. These cosine terms become sizeable only for highly transmitting
channels, and lead in particular to a dc current at finite voltage. This
perfect voltage-bias, non-adiabatic theory, explains quantitatively
\cite{scheer}, in terms of multiple Andreev reflections (MAR) \cite{KBT}, the
strong current non-linearities known as the ``subgap structure'' observed at
finite voltage in all kinds of SNS structures.

\begin{figure}[hptb]
\begin{center}
\includegraphics[width=3.in]{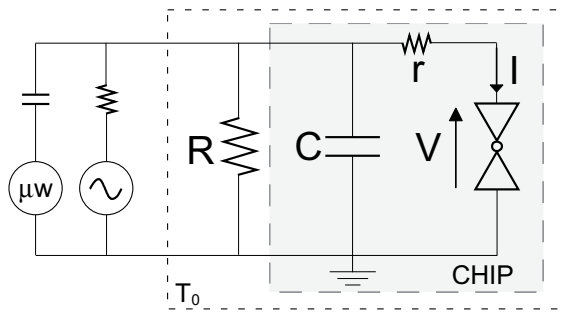}
\caption{ Simplified schematics of the electromagnetic environment
seen by the atomic contact (double-triangle symbol). Dashed grayed
box shows the microfabricated on-chip environment. Dotted box shows
components cooled down to base refrigerator temperature $T_{0}$ . A
low-frequency voltage source coupled through a large resistor
provides a low-frequency current bias. Microwaves are injected
through a small coupling capacitor.\label{circuit}}
\end{center}
\end{figure}

 In the experiments presented here, we monitor the modifications under microwave
irradiation of the dc current-voltage characteristics of voltage biased
 aluminum atomic contacts obtained using microfabricated break junctions
\cite{MicroBJ}. The principle of the experimental setup is shown
schematically in Fig.~\ref{circuit}.
The break junctions are embedded into a biasing circuit of low
impedance (the so-called ``environmental impedance") designed to
approach the perfect voltage bias condition assumed in the theory.
The contact is characterized by its critical current $I_{0}\left(
\left\{ \tau _{i}\right\} \right) $, typically a few tens of nA. It
is placed in series with a microfabricated resistor $r$ and this
combination is shunted by a microfabricated capacitor $C$ and a
surface mounted resistor $R$. In practice we have used two setups,
which differ essentially in the way the current through the atomic
contact is measured. In the first type (A), the current is measured
by means of an array of 100 dc SQUIDs \cite{squid}, as described in
\cite{Steinbach2001}. In the second type (B), the current is
obtained through the voltage drop across the resistor $r$, directly
measured with low-noise voltage amplifiers \cite{samples}. Details
can be found in \cite{pvstein}. Both setups give essentially the
same results.

\begin{figure}[hptb]
\begin{center}
\includegraphics[width=3.in]{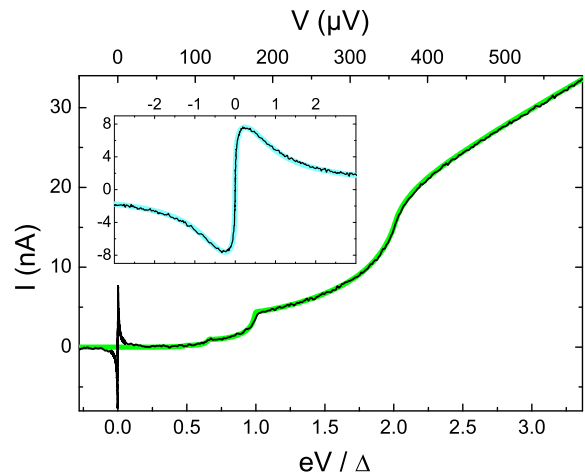}
\caption{(color online). Full lines: current-voltage characteristic
of an Al atomic contact measured in type B setup at refrigerator
temperature $T_{0}=20~\mathrm{mK}$. Grayed line: best fit using zero
temperature MAR theory, obtained for PIN $\left\{ 0.389, 0.238,
0.055\right\} $ and $\Delta=178.2~\mu\mathrm{eV}$. Inset: zoom on
the supercurrent peak. Grayed line: best fit using environment
temperature $T_{e}=133~\mathrm{mK}$ in phase diffusion
theory.\label{supercurrent peak}}
\end{center}
\end{figure}

A typical $I(V)$ of a one atom aluminum contact, with no applied microwaves, is
shown in Fig.~\ref{supercurrent peak}. The strong non-linearities arising at
the thresholds $V=2\Delta /ne$ of the different MAR processes allow to
determine the gap $\Delta $ \cite{gap} and the full $\left\{ \tau _{i}\right\}
$ \cite{scheer}. At small scale (inset of Fig.~\ref{supercurrent peak}), the dc
Josephson current manifests itself as a peak with a finite width. This physics
is well-understood as due to phase fluctuations caused by the noise in the
environmental impedance supposed to be at a finite temperature $T_{e}$. The
theory \cite{Ivanchenko}, developed initially for a purely resistive
environment, is based on the solution of a Langevin equation for the dynamics
of the phase, which diffuses along the Josephson potential. It has been
thoroughly checked experimentally in the case of tunnel junctions
\cite{Steinbach2001}, and for structures containing only channels of small or
intermediate transmission \cite{Goffman}.
 Note that this is an adiabatic
theory, as it does not include the effect of Landau-Zener transitions to the
excited Andreev levels that are essential to explain the experimental results
in the case of highly transmitting channels \cite{Goffman}. The theory has been
extended to the case of structures containing
 ballistic channels \cite{averinBardasIman} and to deal with more general
environments \cite{Duprat2005}. The important fact is that the size of the
supercurrent peak diminishes quite rapidly with the ratio between the Josephson
energy $\varphi _{0}I_{0}$  and the thermal energy $ k_{B}T_{e}$ available in
the dissipative elements of the environment. As shown in the inset of
Fig.~\ref{supercurrent peak}, the agreement between the experimental data and
the calculated phase diffusion curves using the independently measured values
of $\left\{ \tau _{i}\right\}
 $, $r, C$ and $R$ is
excellent. However, in the present experiments the environment temperature
extracted through this analysis was always significantly above that of the
refrigerator \cite{noise}.

\begin{figure}[hptb]
\begin{center}
\includegraphics[width=3.in]{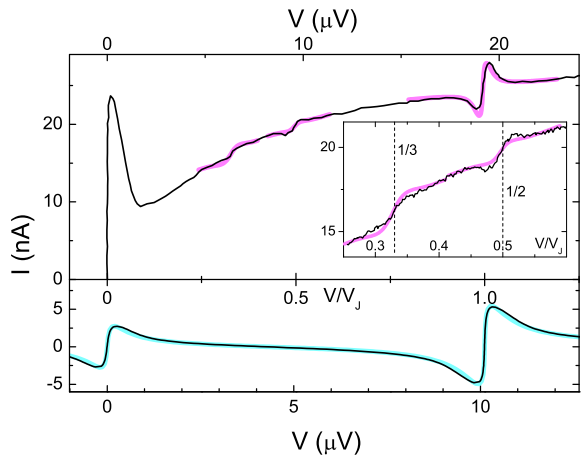}
\caption{(color online). Full lines: $I(V)$s measured  under
microwave excitation for two different contacts, refrigerator
temperature $T_{0}=20~\mathrm{mK}$. Middle axis: voltage in
Josephson voltage units ($V_{J}={\varphi _{0}\omega}$). Upper and
bottom axis: voltage in $\mu\mathrm{V}$. Upper panel: contact on a
type A sample, PIN $\left\{ 0.992, 0.279, 0.278\right\}$,
$\Delta=177~\mu\mathrm{eV}$, $\alpha=0.43$,
$\omega/2\pi=9.3156~\mathrm{GHz}$. Inset: zoom on the small Shapiro
resonances at $V/V_{J}=1/3,1/2$. Grayed lines, predictions of the
mapping model with
 temperature $T_{e}=200~\mathrm{mK}$. Lower curve: Same sample as in
Fig.~\ref{supercurrent peak}, but different run and contact with PIN
$\left\{ 0.573, 0.233, 0.037\right\}$,
$\Delta=179.7~\mu\mathrm{eV}$, $\alpha=0.86$,
$\omega/2\pi=4.892~\mathrm{GHz}$. Grayed line: phase diffusion
theory \cite{Duprat2005} with environment temperature
$T_{e}=120~\mathrm{mK}$.\label{resonances}}
\end{center}
\end{figure}

When microwaves are applied the whole $I(V)$ is deeply modified. As shown in
Fig.~\ref{resonances}, sharp resonances appear at well defined voltages which
scale with the microwave frequency $\omega.$ The amplitude of these resonances,
and of the supercurrent peak itself, oscillates with the amplitude of the
microwave field.
 The non-adiabatic theory \cite{MAR}  has been extended \cite{Cuevas2002}
to consider a perfect voltage bias containing both a constant component $V$ and
an oscillating one $A\cos \left( \omega t\right) $, in which case the phase evolves in time according to $%
\delta \left( t\right) =\omega _{J}t+2\alpha \sin \left( \omega t\right) $ where $\alpha = A/\left( 2\varphi _{0}\omega\right) $%
. The time-dependent current becomes \cite{Cuevas2002}%
\begin{equation}
I(V,\alpha ,\tau ,\omega ,t )=\sum\limits_{m,n} I_{m}^{n}(V,\tau ,\alpha
,\omega)\mathrm{e}^{i \left[ m\omega _{J} +n\omega \right]t }.
\label{currentmicrowaves}
\end{equation}
\begin{figure}[hptb]
\begin{center}
\includegraphics[width=3.in] {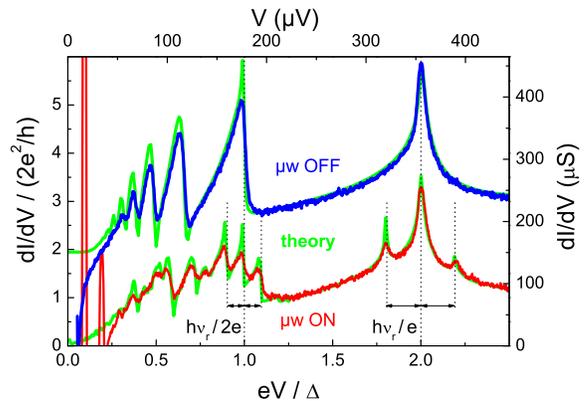}
\caption{(color online). Full lines: measured differential
conductance of contact
 with PIN $\left\{ 0.696, 0.270, 0.076\right\} $
and $\Delta=177.6~\mu\mathrm{eV}$. Upper curve (shifted upwards by
$150~\mu\mathrm{S}$): no microwaves. Lower curve: under microwave
irradiation with $\alpha=0.70$, $\omega/2\pi=8.2935~\mathrm{GHz}$.
Grayed lines: predictions of PAMAR theory, with no adjustable
parameters. The theory includes neither the negative contribution of
the Josephson peak at low voltages, nor the Shapiro resonances.
 \label{pamar}}
\end{center}
\end{figure}

In this case of perfect voltage bias, the dc component can be explicitly
decomposed into a continuous background $I_0^0(V,\tau ,\alpha ,\omega)$ (which
for $\alpha=0$ corresponds to the MAR current), plus a sum of singularities
$I_{m}^{n}(V,\tau ,\alpha ,\omega)\delta \left( V-\frac{n}{m}\varphi
_{0}\omega\right)$ which correspond to the well known Shapiro resonances
\cite{Shapiro} arising from the beatings between the Josephson ac currents and
the external microwave
probe when their frequencies are commensurate ($m\omega _{J}=n\omega$%
). For a contact containing only low and intermediate transmitting channels
(all $\tau $'s$<0.5$), the predicted current-phase relation is almost
sinusoidal and the $m=1$ component is the only sizeable one in the
supercurrent. Therefore, like in the well-known case of tunnel junctions,
Shapiro resonances appear centered at integer multiples of the Josephson
voltage $V _{J}=\varphi _{0}\omega$ determined by the external frequency, as
shown in the lower panel of Fig.~\ref{resonances}.
 Obviously, the shape of the resonances cannot be understood within the constant bias theory \cite%
{Cuevas2002} which does not allow for phase fluctuations. However,
as shown by the underlying grayed line in Fig.~\ref{resonances}, the
shape and size of these resonances can be perfectly accounted for
using the theory by Duprat and Levy Yeyati \cite{Duprat2005} who
have extended the Fokker-Planck treatment of
 \cite{Ivanchenko} to include the microwave drive. For these small transmissions
there is essentially no MAR current in the voltage range of the Shapiro
resonances, and this
adiabatic theory works well. For small transmissions the amplitude of the $n$%
-resonance varies with the reduced microwave probe amplitude $\alpha $
basically as a Bessel function of order $n$ (data not shown), which allows for
a calibration of the microwave driving current. In the top panel of
Fig.~\ref{resonances} we show the results on a contact containing a highly
transmitting channel. The most important qualitative fact is the appearance of
small resonances at fractional multiples of the Josephson voltage. These
so-called fractional Shapiro resonances are a direct consequence of the
deviation of the current-phase relationship from a pure sine function. The
resonances occur at voltages for which there is an important MAR current, which
is itself modified by the microwave field, and the current cannot be decomposed
into two distinct contributions as before. As there exists no theory dealing
with this situation of non-adiabatic phase diffusion in presence of microwaves,
we have developed an empirical model in which the resonances are viewed as
replicas of the supercurrent peak. We take into account the effects of the
environment by mapping the dynamics of the phase around each resonance into the
dynamics around zero voltage in absence of microwaves. In other words, we
suppose that the phase fluctuates around the deterministic dynamics imposed by
an hypothetical perfect voltage bias (both dc and microwave). Under this
hypothesis, the system is governed by a Langevin equation similar to the one
describing the dynamics in absence of microwaves, differing simply by an offset
in voltage $\frac{n}{m}V_{J}\ $ and the following scaling of the parameters:
For each $\frac{n}{m}$ resonance, the Josephson critical current is replaced by
its maximum amplitude predicted by the perfect bias, non-adiabatic theory
\cite{Cuevas2002} for the measured $\left\{ \tau _{i}\right\}
 $, and most importantly, the environment
temperature $T_{e}$ has to be replaced by an effective temperature
$mT_{e}$ \cite{pvstein}. This means that fractional Shapiro
resonances are very rapidly washed out by thermal fluctuations
\cite{Dubos}, as compared to the integer resonances. The underlying
grayed lines of the upper panel in Fig.~\ref{resonances} are the
predictions of this mapping approach, the environment temperature
being the only adjustable parameter. The best fit to the data is
obtained assuming an environment temperature of
$T_{e}=200~\mathrm{mK}$ instead of the actual temperature read by
the thermometers $T_{0}=20~\mathrm{mK}$. A linear background term
has also been added to account, at least partially, for the
background current on which the Shapiro resonances superimpose. The
model describes the general trends of the experimental results. In
particular the amplitude of the integer resonances as a function of
the amplitude of the microwave field are quite well accounted for
(data not shown). However, the amplitude of the fractional
resonances is too small to make a quantitative comparison with
theory.

For the typical microwave frequencies $(<12~\mathrm{GHz})$ and amplitudes used
here, the Shapiro resonances are observed over a small voltage range
$(|eV|<0.2\Delta)$. However, at larger voltages there is still a large effect
of the irradiation on the $I(V)$. Figure~\ref{pamar} shows a comparison of the
measured and calculated differential conductance $dI/dV$, in presence of
microwaves. With no microwaves, the onsets of the different MAR processes of
Fig.~\ref{supercurrent peak} appear as peaks on the differential conductance
curve. In presence of microwaves, satellite peaks appear around them, at
voltages $V=(2\Delta \pm m\hbar \omega _{r})/2ne $. They correspond to the
absorption or emission of $m$ photons during the MAR process which transfers
$n$ electronic charges, i.e. to photon-assisted MAR processes (PAMAR). The
experimental results are very well reproduced by the dc component $I_0^0(V,\tau
,\alpha ,\omega)$ of Eq. \ref{currentmicrowaves}, with no adjustable
parameters. Note that the theory does not consider the effect of thermal
fluctuations of the phase. For all the contacts we have measured, the agreement
between theory and experiment is as good as shown in Fig.~\ref{pamar}. Although
these multiphoton processes have been already observed \cite{Dayem} and
identified \cite{TienGordon}, to our knowledge this is the first direct
quantitative comparison between theory and experiment.

In conclusion, the ability of tuning and measuring the transmission of the few
channels accommodated by atomic contacts allows to compare, with no adjustable
parameters, experimental results with the predictions of the modern theory of
the Josephson effect. The observation of fractional Shapiro resonances is clear
indication of the occurrence of supercurrents at harmonics of the Josephson
frequency in contacts of large transmission. Furthermore, we find quantitative
agreement between our results and the predictions of the theory of
photon-assisted multiple Andreev reflections. The results illustrate the power
of this modern view, which is able to describe both dissipative and
non-dissipative currents, simply in terms of occupation of Andreev levels.

\begin{acknowledgments}
This work was supported by the EU Network DIENOW. We acknowledge technical
assistance of P.F. Orfila and P. Senat, and important discussions with J.\ C.
Cuevas, R. Duprat, G. Rubio Bollinger, A. Mart\'{\i}n Rodero and A. Levy
Yeyati, who also provided us with their computer codes.
\end{acknowledgments}


\begin{thebibliography}{}
\bibitem{Fourier} Under voltage bias the Andreev levels acquire a finite width due to
the coupling with the continuum of states beyond the gap, and the theory only
provides $I_{m}(V,\tau )$ as integrals that have to be evaluated numerically
\cite{MAR}.
\bibitem{samples} Both types of samples are
fabricated on metallic substrates(ground plane). The break junctions are
tailored in a $100~\mathrm{nm}$ thick Al film. In samples of type A, the
on-chip
 resistor $r\thicksim7.5~\mathrm{\Omega}$ is a $165 ~\mu \mathrm{m}$-long, $28~\mu
 \mathrm{m}$-wide, $60~\mathrm{nm}$-thick Au:Cu (3:1 in weight) film.
  The capacitor $C\thicksim100~\mathrm{pF}$ is formed between a $4~\mathrm{mm}^{2}$ square of
 the same film and the substrate. The dielectric is a $1.6 ~\mu \mathrm{m}$-thick polyimide
layer. Shunt was $R\thicksim30~\mathrm{\Omega}$. In samples of type B, designed
for four-point measurements, there are four identical copies of each
microfabricated element. The $70 ~\mu \mathrm{m}$-long, $0.5 ~\mu
\mathrm{m}$-wide on-chip resistors $r\thicksim40 ~\mathrm{\Omega}$ are tailored
in a $30~\mathrm{nm}$-thick Au film. The top plates of the capacitors
$C\thicksim92~\mathrm{pF}$ are $200~\mathrm{nm} $-thick Au electrodes. The
shunt resistor was $R\thicksim50~\mathrm{\Omega}$. Values of resistances and
capacitances where measured independently and varied within $1\%$ from sample
to sample.
\bibitem{noise} Both the dc and the ac bias can slightly heat up the on-chip dissipative elements,
but in the present case the excess temperature is mainly due to
noise (not necessarily thermal) originating in the amplifiers and
the biasing sources and which we couldn't completely filter out.
This was particulary important for the type A setup.
\end{thebibliography}
\end{document}